\documentstyle[12pt]{article}

\def\a{\alpha}
\def\b{\beta}
\def\m{\mu}
\def\tm{\tilde{m}}
\def\th{\theta}
\def\tr{{\rm Tr}}
\def\pf{{\rm Pf}}

\def\P{\Phi}

\def\l{\lambda}
\def\L{\Lambda}

\def\e{\epsilon}

\begin{document}

\newcommand{\inv}[1]{{#1}^{-1}} 

\renewcommand{\theequation}{\arabic{equation}}
\newcommand{\beq}{\begin{equation}}
\newcommand{\eeq}[1]{\label{#1}\end{equation}}
\newcommand{\ber}{\begin{eqnarray}}
\newcommand{\eer}[1]{\label{#1}\end{eqnarray}}
\begin{center}
     April, 1995
                                \hfill    RI-4-95\\
                                \hfill    hep-th/9504080\\

\vskip .3in

{\large \bf More Results in $N=1$ Supersymmetric Gauge Theories }
\vskip .4in

{\bf S. Elitzur}, \footnotemark\
\footnotetext{e-mail address: elitzur@vms.huji.ac.il}
{\bf A. Forge}, \footnotemark\
\footnotetext{e-mail address: forge@vms.huji.ac.il}
{\bf A. Giveon}, \footnotemark\
\footnotetext{e-mail address: giveon@vms.huji.ac.il}
{\bf E. Rabinovici} \footnotemark\
\footnotetext{e-mail address: eliezer@vms.huji.ac.il}
\vskip .1in

{\em Racah Institute of Physics, The Hebrew University\\
  Jerusalem, 91904, Israel} \\

\vskip .1in
\end{center}
\vskip .2in
\begin{center} {\bf ABSTRACT } \end{center}
\begin{quotation}\noindent

We present the exact effective superpotentials in $4d$, $N=1$
supersymmetric $SU(2)$ gauge theories with $N_3$ triplets and $N_2$
doublets of matter superfields.
For the theories with a single triplet matter
superfield we present the exact gauge couplings for
{\em arbitrary} bare masses and Yukawa couplings.

\end{quotation}
\vfill
\eject
\def\baselinestretch{1.2}
\baselineskip 16 pt

\noindent

Recently, many new exact results were derived in four
dimensional supersymmetric field theories (for a review, see ref.
\cite{S}).
In this note we report the results~\footnote{
A more detailed derivation of the results will be furnished in
\cite{efgr}.} of applying the methods of refs.
\cite{S,ILS,I,IS1} to the general case of an infra-red non-trivial
$N=1$ supersymmetric gauge theory with an $SU(2)$
gauge group, $N_3$ matter supermultiplets in the adjoint
representation, $\P_{\a}^{ab}$, $\a=1,...,N_3$,
and $N_2=2N_f$ supermultiplets in the fundamental representation,
$Q_i^a$, $i=1,...,2N_f$. Here $a,b$ are fundamental representation indices,
and $\P^{ab}=\P^{ba}$. To preserve asymptotic freedom or
conformal invariance, we need to impose negative or vanishing beta
functions; it implies the necessary condition:
\beq
b_1=6-N_f-2N_3\geq 0,
\eeq{b1}
where $-b_1$ is the one-loop coefficient of the gauge coupling
beta-function.
We consider these models in the presence of Yukawa couplings, $\l$, and masses
$m$. The effective potential is obtained by what is called the
``integrating in'' method \cite{ILS,I}. Under certain conditions, one may,
unconventionally, derive the effective superpotential for modes which are of
finite mass, given the effective action in which the modes have been
considered to have infinite mass. We apply the integrating in technique when
it is valid \cite{I}; the various consistency checks to
which the result is subjected strengthen the reliability of the method.
We integrate in matter in the adjoint representation given the exact
effective action for $N_3=0$ and $2N_f$ doublets\footnote{
The $N_3=0$, $2N_f$ superpotential can also be derived
by integrating in doublets
to the pure  $SU(2)$, $N=1$ supersymmetric Yang-Mills theory \cite{ILS,I}.}.
We obtain the superpotential\footnote{
When $b_1=4$ one also obtains constraints; they will be discussed soon.
When $b_1=0$, ``$\L^{-b_1}$'' in (\ref{W}) should be replaced by a function
of $\tau_0={\th_0\over\pi}+{8\pi i\over g_0^2}$ (the non-Abelian gauge
coupling) and $\det \l$.}
\ber
W_{N_f,N_3}(M,X,Z) &=& -(4-b_1)\Big\{\L^{-b_1} \pf X\Big[
{\rm det}_{N_3}(\Gamma_{\a\b})\Big]^2
\Big\}^{1/(4-b_1)}\nonumber\\
&+&\tr_{N_3} \tm M +{1\over 2}\tr_{N_2} mX
+{1\over\sqrt{2}}\tr_{N_2} \l^{\a} Z_{\a} ,
\eer{W}
where
\beq
\Gamma_{\a\b}(M,X,Z)=M_{\a\b}+\tr_{N_2}(Z_{\a}X^{-1}Z_{\b}X^{-1}).
\eeq{G}
Here $\L$ is the dynamically generated scale, while $\tm_{\a\b}$, $m_{ij}$
and $\l^{\a}_{ij}$ are the bare masses and Yukawa couplings, respectively
($\tm_{\a\b}=\tm_{\b\a}$, $m_{ij}=-m_{ji}$,
$\l^{\a}_{ij}=\l^{\a}_{ji}$). The gauge singlets, $X$, $M$, $Z$, are
given in terms of the $N=1$ superfield doublets, $Q^a$, and the triplets
$\P^{ab}$,  as follows:
\ber
X_{ij}&=&\e_{ab}Q_i^a Q_j^b, \qquad  a,b=1,2, \qquad i,j=1,...,N_2=2N_f,
\nonumber\\
M_{\a\b}&=&\e_{aa'}\e_{bb'}\P_{\a}^{ab}\P_{\b}^{a'b'}, \qquad \a ,\b=1,...,N_3,
\nonumber\\
Z_{ij}^{\a}&=&\e_{aa'}\e_{bb'}Q_i^a\P_{\a}^{a'b'}Q_j^b.
\eer{XMZ}
{}From eq. (\ref{XMZ}) it is clear that the determinant in $W_{N_f,N_3}$
vanishes classically, namely:
$\Gamma_{\a\b}(M,X,Z)=0$ is a classical constraint.
Quantum mechanically, the constraint is removed; by taking the $\L\to 0$
limit in eq. (\ref{W}), one recovers the classical constraint
${\rm det}_{N_3}(\Gamma_{\a\b})=0$ (if $b_1<4$).

The first part of $W$ in (\ref{W}) is the main result of this paper; it is
the exact non-perturbative superpotential. The superpotential is expressed
in terms of particular combinations of the gauge singlets $M,X,Z$. Among
other things, it contains the information necessary to derive various
subsequent results in this paper.

We consider the general case for the masses of the superfields $Q^a$.
They become massless if $m=\l=0$. In
case the doublets are massive, one can obtain the low-energy effective
action for the superfields $M$ by integrating out the singlets $X$ and $Z$
from $W$.

\vskip .1in

Models without triplets ($N_3=0$) were studied in \cite{study,S1}.
The superpotential is
\beq
W_{N_f,0}(X)=(2-N_f)\L^{{6-N_f\over 2-N_f}} (\pf X)^{{1\over N_f-2}}+{1\over
2}\tr_{N_2} mX .
\eeq{WX}
For $N_f=1$, the massless superpotential reads:
$W=\L^5/X$. For $N_f=2$ ($b_1=4$ in eq. (\ref{b1})), $W=0$, and by
the integrating in procedure we also get the constraint: $\pf X=\L^4$. For
$N_f>2$, $W$ is proportional to some positive power of the classical
constraint: $\pf X=0$.

Models without doublets ($N_f=0$) were studied in \cite{SW1,IS1,IS2}.
In these cases
\beq
W_{0,N_3}(M)=
2(1-N_3)\L^{{N_3-3\over N_3-1}}(\det M)^{{1\over N_3-1}}+\tr_{N_3}\tm M.
\eeq{WM}
The massless $N_3=1$ case is a pure $SU(2)$,
$N=2$ supersymmetric Yang-Mills theory.
This model was considered in detail in ref. \cite{SW1}. In this case, $W=0$
(compatible with eq. (\ref{W})). As in the other $b_1=4$ case, discussed
above, by the integrating in procedure one also gets a constraint in this
case: $M=\pm \L^2$. This result can be understood as the starting point of
the integrating in procedure is a pure $N=1$ supersymmetric
Yang-Mills theory. Therefore, it
leads us to the points at the edge of confinement in the moduli space.
These are the two singular points in the
$M$ moduli space of the theory; they are due to massless monopoles or
dyons. Such excitations are not constructed out of the elementary
degrees of freedom and, therefore, there is no trace for them in $W$. (This
situation is different if $N_f\neq 0$; in this case, monopoles are
different manifestations of the elementary degrees of freedom.)

The $N_f=0$, $N_3=2$ case is discussed in refs. \cite{IS1,IS2}. In this
case,
the superpotential in eq. (\ref{W}) is the one presented in \cite{IS1,IS2} on
the confining and the oblique confinement branches. As in the $N_3=1$ case,
this is because the starting point of
the integrating in procedure is a pure $N=1$ supersymmetric
Yang-Mills theory and, therefore, it
leads us to the confining branches in the moduli space.

For $N_3=3$ there is an
additional Yukawa coupling that we did not consider in (\ref{XMZ}):
the one which couples the three (antisymmetric) triplets.
Therefore, we should also integrate in the
additional gauge singlet $\P\P\P\equiv \det\P$. The superpotential in
eq. (\ref{WM}) remains valid also in the presence of
$W_{{\rm tree}}=\l\det\P$ because $\det\P=(\det M)^{1/2}$; the Yukawa
coupling, $\l$, replaces ``$\L^0$'' in eq. (\ref{WM}).
This result coincides with the one derived in \cite{IS2}.
In the massless case, this
theory flows to an $N=4$ supersymmetric Yang-Mills fixed point.

\vskip .1in

In the rest of this note we consider the models with a single adjoint
matter: $N_3=1$, and with fundamental matter: $N_f\neq 0$.
In this case $M$ is a complex modulus. All these models have a coulomb
phase and thus an effective Abelian gauge field coupling,
$\tau(M,m,\l)=\th/\pi+8\pi i/g^2$, can be defined for them. The complexified
gauge coupling depends on the
modulus superfield, as well as the bare masses and Yukawa couplings.
The quantum theory is invariant under the $SL(2,Z)$ duality transformations
acting on $\tau$ \cite{CR} and,
therefore, it is convenient to define $\tau(M,m,\l)$ by the elliptic curve
equation:
\beq
y^2=x^3+a(M,m,\l)x^2+b(M,m,\l)x+c(M,m,\l).
\eeq{yx}

We can use the superpotential $W_{N_f,1}$ in (\ref{W}) in order to find
$\tau$. This is done as follows. As was mentioned before, for $N_f>0$, all
the degrees of freedom
that may become massless somewhere in the $M$ moduli space are already
present in the superpotential. Therefore, the solutions to the equations of
motion, derived from $W$ by variations with respect to $X$ and $Z$,
must coincide with the singularities of the elliptic curve (\ref{yx}). This
is because for values of $M$ which extremize $W$, some charged massive modes
become massless, and thus give rise to these singularities. In this way we can
derive the coefficients $a,b,c$ from $W$. For $N_f=1$ this was already done
in ref. \cite{IS1}; one finds
\beq
a=-M, \qquad b={\L^3\over 4} m, \qquad c=-{\a\over 16},
\eeq{abc1}
where
\beq
\a\equiv {\L^{2b_1}\over 2^{2N_f}} \det\l = {\L^6\over 4} \det\l .
\eeq{aL}

For $N_f=2$ one finds
\ber
a&=& -M, \qquad b\,\,=\,\, -{\a\over 4}+{\L^2\over 4}\pf m , \nonumber\\
c&=& {\a\over 8}\Big(2M+\tr(\m^2)\Big),
\eer{abc}
where
\beq
\a\equiv {\L^{2b_1}\over 2^{2N_f}} \det\l = {\L^4\over 16} \det\l, \qquad
\m=\l^{-1}m.
\eeq{aL2}
Equation (\ref{abc}) generalizes the result of ref. \cite{SW2} to {\em
arbitrary} bare masses and Yukawa couplings.
Indeed, in the $N=2$ supersymmetric case (namely, when $\l={\rm
diag(\l_1,\l_2)}$, where $\l_1,\l_2$ are $2\times 2$ matrices with
$\det\l_1=\det\l_2=1$, and $m={\rm diag}(m_1\e,m_2\e)$, where $\e$ is the
standard $2\times 2$ constant antisymmetric matrix), the result (\ref{abc})
coincides with the one obtained in ref. \cite{SW2}.
All the symmetries and quantum numbers of the various parameters, as used
in \cite{SW1,SW2}, are already embodied in the superpotential $W$ of eq.
(\ref{W}).

We have also used the same procedure to derive the elliptic curves in the
$N_f=3,4$; $N_3=1$ cases, for {\em arbitrary} Yukawa couplings and masses.
For $N_f=3$ one finds
\ber
a&=& -M-\a ,   \nonumber\\
b&=& 2\a  M + {\a\over 2}\tr(\m^2) + {\L\over 4} \pf m ,
\nonumber\\
c&=&{\a\over 8}\Big( -8M^2-4M \tr(\m^2) - [\tr(\m^2)]^2 + 2\tr(\m^4)
\Big) ,
\eer{abc3}
where
\beq
\a\equiv {\L^{2b_1}\over 2^{2N_f}}\det\l={\L^2\over 64}\det \l,
\qquad \m=\l^{-1}m.
\eeq{am}
For $N_f=4$ one finds
\ber
a&=&{1\over \b^2}\Big\{
2{\a+1\over \a-1}M+{8\over \b^2}{\a\over (\a-1)^2}\tr(\m^2)\Big\}, \nonumber\\
b&=&{1\over \b^4}\Big\{
-16{\a\over (\a-1)^2} M^2 + {32\over \b^2}{\a(\a+1)\over (\a-1)^3}M\tr(\m^2)
\nonumber\\
&-&{8\over \b^4}{\a\over (\a-1)^2}\Big[(\tr (\m^2))^2-2\tr(\m^4)\Big]+
{4\over \b^4}{(\a+1)\L^{b_1}\over (\a-1)^2} \pf m\Big\} , \nonumber\\
c&=& {1\over \b^6}\Big\{
-32{\a(\a+1)\over (\a-1)^3}M^3+{32\over \b^2}{\a(\a+1)^2\over
(\a-1)^4}M^2\tr(\m^2)\nonumber\\
&+&M\Big[-{16\over \b^4}{\a(\a+1)\over (\a-1)^3}
\Big((\tr(\m^2))^2-2\tr(\m^4)\Big) + {32\over \b^4}{\a\L^{b_1}\over (\a-1)^3}
\pf m \Big]\nonumber\\
&-&{32\over \b^6}{\a\over (\a-1)^2}\Big[\tr(\m^2)\tr(\m^4)-{1\over
6}(\tr(\m^2))^3-{4\over 3}\tr(\m^6)\Big]\Big\}.
\eer{abc4}
Here $\a$ and $\b$ are functions of $\tau_0$, the non-Abelian
gauge coupling constant; comparison with ref. \cite{SW2} gives
\beq
\a(\tau_0)\equiv {``\L^{2b_1}"\over 2^{2N_f}} \det\l
=\left({\th_2^2-\th_3^2\over \th_2^2+\th_3^2}\right)^2, \qquad
\b(\tau_0)={\sqrt{2}\over \th_2\th_3}, \qquad \m=\l^{-1}m,
\eeq{g2}
where
\beq
\th_2(\tau_0)=\sum_{n\in Z}(-1)^n e^{\pi i \tau_0 n^2}, \qquad
\th_3(\tau_0)=\sum_{n\in Z}e^{\pi i \tau_0 n^2}, \qquad
\tau_0={\th_0\over \pi}+{8\pi i\over g_0^2}.
\eeq{gth}
($16\a^{1/2}(\det\l)^{-1/2}$ replaces ``$\L^{b_1}$'' in eq. (\ref{W});
$\a(\tau_0)$ is dimensionless, and has zero $U(1)_R\times U(1)_Q\times
U(1)_{\P}$ quantum numbers.)\footnote{
$(\det\l)^{-1/2}$ has the correct quantum numbers needed for the matching
condition, $\a^{1/2}(\det\l)^{-1/2}\tm=\L_{N_f=4,N_3=0}$, which
we used in the integrating in procedure.
To compare eq. (\ref{abc4}) with ref. \cite{SW2} we need to take $m={\rm
diag}(m_1\e,m_2\e,m_3\e,m_4\e)$ and
$\l={\rm diag}(\l_1,\l_2,\l_3,\l_4)$, where $\l_I$, $I=1,2,3,4$, are
$2\times 2$ matrices with $\det(\l_I)=1$. In this case,
$\tr(\m^2)=-2\sum_{I=1}^4 m_I^2$,
$(\tr(\m^2))^2-2\tr(\m^4)=8\sum_{I<J}m_I^2 m_J^2$,
$\tr(\m^2)\tr(\m^4)-{1\over 6}(\tr(\m^2))^3-{4\over
3}\tr(\m^6)=8\sum_{I<J<K} m_I^2 m_J^2 m_K^2$.}

In eq. (\ref{abc4}) $M$ is rescaled with respect to $M$ in the
superpotential, $M\to \b^2 M$, using the scale invariance of the $N=2$ theory
with four flavors~\footnote{
Note that $W$ in eq. (\ref{W}) scales appropriately, $W\to \b^3 W$, under
the scale transformation: $\P\to \b\P, Q\to \b Q, m\to\b m, \L\to \b\L,
\l\to\l$.}.

The $S$-duality symmetry is valid in the $N_3=1, N_f=4$ theories for {\em
arbitrary} $\l, m$,  similar to the $SL(2,Z)$ invariance in the presence of
masses discussed in
ref. \cite{SW2}. The $SL(2,Z)$ transformations map $\tau_0$ to
$(a\tau_0+b)(c\tau_0+d)^{-1}$, $a,b,c,d\in Z$, $ad-bc=1$. Combined
with triality (which acts on $\m$), it leaves the elliptic curve invariant.

Equations (\ref{abc3}, \ref{abc4}) generalize the results
obtained in \cite{SW2} for the $N=2$
supersymmetric case (namely, for an appropriate subspace of $m$ and $\l$).
Needless to say that by taking the mass $m_{N_2,N_2-1}$ to infinity, we can
generate the effective superpotential with $N_f-1$ flavors from the solution
with $N_f$ flavors (by integrating out), as well as the corresponding
elliptic curve.

\vskip .1in

We shall end with a few remarks:

\begin{itemize}

\item
The derivation of the elliptic curves from the
superpotentials, in all $N_3=1$ cases ($N_f=1,2,3,4$ in eqs. (\ref{abc1}),
(\ref{abc}), (\ref{abc3}), (\ref{abc4})), suggests that the variable $x$ in
eq. (\ref{yx}) could be identified with $\Gamma$ in eq. (\ref{G})
(up to a shift by $M$).

\item
The techniques used, and the patterns uncovered in this note can be applied
also to the $N_3>1$ cases
(with mass given to part of the $M$ fields), and to other gauge groups. We
shall report on that in \cite{efgr}.

\item
The $SU(2)$, $N_3=1$, $N_f$ models fall into a lacuna in the analysis in
ref. \cite{K} of the dual models to $SU(N_c)$ systems with matter in the
adjoint and fundamental representations. The results obtained here might shed
some light on this gap~\footnote{We thank N. Seiberg for a discussion
on this point.}.

\item
Finally, to complete the survey of models obeying eq. (\ref{b1}), let us note
that one can also have an infra-red non-trivial theory
with a single matter superfield in the $I=3/2$ representation ($N_4=1$ in
our notation). The $N_4=1$, $N_f=0$ theory was shown to have $W=0$
\cite{ISS}. Adding $N_f=1$ matter results with $b_1=0$ in eq. (\ref{b1}).
The two-loop beta function renders the theory infra-red free. As no Yukawa
coupling is possible, this model is indeed infra-red free.

\end{itemize}

\vskip .3in \noindent
{\bf Acknowledgements} \vskip .2in \noindent
We thank S. Forste and N. Seiberg for discussions.
The work of SE is supported in part by the BRF - the Basic Research
Foundation.
The work of AG is supported in part by BSF - American-Israel Bi-National
Science Foundation, by the BRF,  and by an Alon fellowship.
The work of ER is supported in part by BSF and by the BRF.

\end{document}